\documentclass[12pt]{article}
\usepackage{amssymb,amsmath,amsthm,graphicx,ulem}

\usepackage{graphicx,subfigure}
\usepackage{epsfig}
\usepackage{amsmath}
\usepackage{amsfonts}
\usepackage{amssymb}
\usepackage[usenames]{color}
\usepackage[letterpaper,left=2.cm,right=2.cm,top=2.5cm,bottom=2.5cm]{geometry}
\usepackage{float}
\usepackage{hyperref}

\newcommand{\beq}{\begin{equation}}
\newcommand{\eeq}{\end{equation}}
\newcommand{\be}{\begin{equation}}
\newcommand{\ee}{\end{equation}}
\newcommand{\beqa}{\begin{eqnarray}}
\newcommand{\eeqa}{\end{eqnarray}}
\newcommand{\beqar}{\begin{eqnarray*}}
\newcommand{\eeqar}{\end{eqnarray*}}
\newcommand{\bea}{\begin{eqnarray}}
\newcommand{\eea}{\end{eqnarray}}

\newcommand{\p}{\partial}

\newcommand{\reef}[1]{(\ref{#1})}

\numberwithin{equation}{section}
\begin{document}
 
\allowdisplaybreaks
\normalem
\title{A No Black Hole Theorem}
\author{Gavin S. Hartnett${}^1$, Gary T. Horowitz${}^1$, and Kengo Maeda${}^2$
\\ \\ 
${}^1$Department of Physics, UCSB, Santa Barbara, CA 93106, USA 
\\
${}^2$Faculty of Engineering, Shibaura Institute of Technology, Saitama 330-8570, JAPAN
\\ 
\small{hartnett@physics.ucsb.edu, gary@physics.ucsb.edu, maeda302@sic.shibaura-it.ac.jp}}

\date{}
\maketitle 

\begin{abstract}
We show that one cannot put a stationary (extended) black hole inside certain gravitating flux-tubes. This includes an electric flux-tube in five-dimensional Einstein-Maxwell theory, as well as the standard flux-branes of string theory. The flux always causes the black hole to grow indefinitely. One finds a similar restriction in a Kaluza-Klein setting where the higher dimensional spacetime contains no matter.
\end{abstract}

\newpage

\baselineskip16pt

\section{Introduction}
Our basic understanding of black holes includes a series of ``no-hair" theorems which state that certain types of matter must vanish outside stationary black holes. Intuitively, this says that the matter either falls into the black hole or radiates out to infinity. These theorems were originally proven for  linear fields, but they were thought to hold more generally. It was later realized that many types of hair are indeed possible. One class of examples consists of black holes in anti-de Sitter space. Perhaps the simplest case is a charged scalar field which can exist outside certain charged black holes \cite{Gubser:2008px,Hartnoll:2008vx}. A large class of asymptotically flat examples involves theories that admit stationary solitons. One can often put a small stationary black hole inside the soliton without destroying it \cite{Kastor:1992qy}. This includes solutions of Einstein-Yang-Mills theory \cite{Bartnik:1988am,Volkov:1989fi,Bizon:1990sr}, and monopoles of the Einstein-Yang-Mills-Higgs system \cite{Lee:1991vy}. (For a comprehensive review of these solutions, see \cite{Volkov:1998cc}.)  One can also put a rotating black hole inside a rotating boson stars \cite{Dias:2011at,Herdeiro:2014goa}.
 
There are some cases where one cannot put a static black hole inside a static soliton. For example, if one puts a small black hole inside a static perfect fluid star it will grow and slowly consume the star.  Similarly, one cannot put a static black hole inside a static boson star\footnote{A subtlety here is that although the metric is static, the scalar field has a time dependent phase and is not static itself.} \cite{Pena:1997cy}. The matter content of the soliton determines whether or not one can put a static black hole inside.

In higher dimensions, there are extended black holes such as black strings and black branes. There are also extended solitons such as flux-branes. These are higher dimensional generalizations of ``Melvin's magnetic universe" \cite{Melvin:1963qx} which describes a static, cylindrically symmetric, gravitating magnetic flux-tube. A natural question is whether one can put a static black brane inside a static flux-brane. Since there is an exact four dimensional static solution describing a Schwarzschild black hole inside Melvin's magnetic universe \cite{Ernst}, one might expect the answer is yes. We will show  that this is incorrect. If one puts a  black brane inside a flux-brane, it will necessarily grow and consume the brane.

The simplest example starts with an electric flux-tube in five dimensions. As this solution does not appear to exist in the literature, we will numerically construct it in Sec.~\ref{sec:fluxbranes}. That is,  we find a static, cylindrically symmetric solution of the five-dimensional Einstein-Maxwell theory describing a self-gravitating electric flux-tube. The product of four-dimensional Schwarzschild and a line is a simple five-dimensional black string with the same symmetry as the flux-tube. Our result shows that one cannot put a thin black string inside the flux-tube and keep the solution stationary.

A much larger class of examples that have been discussed in the literature are the flux-branes of string theory \cite{Gutperle:2001mb}, which are sourced by one of the higher rank forms in supergravity. We will review these solutions in Section 4. Our result implies that one cannot put a stationary black brane inside any of these flux-branes \footnote{It was earlier shown \cite{Emparan:2010ni} that certain rank forms cannot exist outside static black holes which are asymptotically flat in all directions.}.
 
As a final application of our result, one can consider higher dimensional vacuum solutions with a $U(1)$ symmetry. Under Kaluza-Klein reduction there is a Maxwell field, and one can construct an electric flux-tube solution of the dimensionally reduced theory. Since one cannot put a stationary black string inside this flux-tube, it follows that one cannot put a black two-brane in the higher dimensional vacuum solution. At first sight, this example is more surprising than the first two. When there is nonzero flux outside the horizon, one can imagine that a stationary black hole cannot exist since the flux falls in causing the black hole to grow. But in a vacuum solution, there is no matter to fall in.

The answer to this puzzle lies in the Raychaudhuri equation applied to the null geodesic generators $\ell^a$ of the event horizon (in $D$ spacetime dimensions):
\be\label{Raychaudhuri}
\frac{d\theta}{d\lambda} = -\frac{1}{D-2}\theta^2 - \sigma_{ab}\sigma^{ab} - T_{ab} \ell^a \ell^b
\ee
where $\lambda$ is an affine parameter, $\theta$ is the expansion, and $\sigma_{ab}$ is the shear of the null geodesic congruence. For a stationary black hole, the left hand side must vanish. Since the theories we are interested in all satisfy the null energy condition, the right hand side is the sum of three negative terms. In order for a stationary black hole to exist, each one much vanish. This is indeed possible for black holes placed inside some solitons \cite{Volkov:1998cc}, but we will see that for the flux-branes, the last term is always nonzero. In the Kaluza-Klein example where the higher dimensional solution has no matter, we will see that the shear is necessarily nonzero on the horizon causing the horizon to grow.

The outline of this paper is as follows. In the next section we prove that a uniform black brane cannot remain stationary inside a flux-brane. In section 3, we discuss the generalization to nonuniform black branes. We will prove that there cannot exist stationary black branes which are nonuniform in a compact direction and argue that similar results hold for noncompact directions also. In section 4 we discuss the flux-branes in detail, and construct some numerically. We conclude in section 5 with some open questions. The two appendices contain some technical details.

\section{No Uniform Black Branes \label{sec:Theorem}}
Consider a theory of gravity in $D$ dimensions coupled to a closed $(p+1)$-form field strength $F_{p+1}$. We will assume a general (Einstein-frame) action of the  form
\be\label{action}
S = \int d^D x \sqrt{-g} \left [ R - \frac{1}{2} (\nabla \phi)^2 - \frac{1}{2(p+1)!} e^{a \phi} F_{p+1}^2 \right]
\ee
Note that we have  included a possible (but not required) scalar field with coupling to $F_{p+1}$ governed by a constant $a$. (One can also include additional matter fields or Chern-Simons terms  and they will not affect our argument.) The equations of motion following from \reef{action} are
\bea \label{eoms}
&& R_{ab} = \tau_{ab}, \qquad  d \star \left(e^{a\phi} F_{p+1} \right) = 0, \qquad \nabla^2 \phi - \frac{a e^{a\phi}}{2(p+1)!}F_{p+1}^2 = 0, 
\eea
where $\tau_{ab}$ is the trace-reversed stress tensor $\tau_{ab} = T_{ab} - \frac{T}{D-2} g_{ab}$ given by
\be \tau_{ab} = \frac{1}{2}\nabla_a \phi \nabla_b \phi + \frac{e^{a\phi}}{2}\left[ \frac{1}{p!} (F_{p+1})_{a...}(F_{p+1})_b{}^{...} - \frac{p}{(D-2)(p+1)!}g_{ab}F_{p+1}^2 \right]. \ee
In addition, one has the constraint  $dF_{p+1} =0$. There are ``flux-brane" solutions to this theory which are nonsingular solutions with at least $p+1$ (commuting) translational symmetries which include time, and $F_{p+1}$ is nonzero when restricted to this homogeneous subspace. We will not need the detailed form of these solutions to prove our ``no black hole theorem" so we delay our construction of these solutions until Sec.~\ref{sec:fluxbranes}. In this section we rule out uniform black branes inside these flux-branes, and in the next section we will argue that similar results hold for the nonuniform case.

\vskip .5cm

\noindent{\bf Theorem:} Consider a ``flux-brane" solution to \reef{action}, i.e., a nonsingular solution  in which $F_{t x_1 \cdots x_p}$ is nonzero, and all fields are independent of $t,x_1, \cdots, x_p$.   Then one cannot put a stationary, translationally invariant, black $p$-brane in the center of this flux-brane.

\vskip .5cm
\noindent{\bf Proof:} First note that $F_{t x_1 \cdots x_p} \equiv E$ must be constant. If it were a function of another coordinate, say $r$, then  $dF = 0$ would require that $F$ has a component $F_{r\cdots}$ that depends on $t$ or one of the $x_i$ contradicting the translation invariance. We will first rule out static black branes, and then generalize to the stationary case. If there were a solution describing a static black $p$-brane in the center of this flux-brane with horizon at $r=r_0$, then $F_{t x_1 \cdots x_p} F^{t x_1 \cdots x_p} \to \infty$ as $r \to r_0$. To obtain a flux which is smooth on the future horizon, one can introduce $F_{rx_1 \cdots x_p}(r)$ so that the $t$ and $r$ components of $F_{p+1}$ combine to give $F_{vx_1 \cdots x_p} = E$, where $v = t + h(r)$ is a good coordinate near the horizon. 
The static Killing field which is null on the horizon is now $\xi =\p/\p v$ (since we have not changed the radial coordinate), so the flux of energy crossing the horizon is

\be\label{flux}
T_{bc} \xi^b \xi^c \propto e^{a\phi} (\xi^b F_{b  \cdots })( \xi^c {F_c}^{  \cdots}) = e^{a\phi} F_{v i \cdots j} {F_v}^{ i \cdots j}
\ee
This cannot vanish since the right hand side is a sum of nonnegative terms with at least one positive contribution coming from $F_{vx_1 \cdots x_p}$.\footnote{The metric on the $x_1, \cdots, x_p$ subspace must be positive definite since the horizon is a null surface, and these coordinates denote directions along a cross section of the  horizon, not along a null generator.} This contradicts the assumption that the black brane was static, since the Raychaudhuri equation \reef{Raychaudhuri} shows that the horizon must grow  when there is an energy flux across the horizon.
  
This argument is easily extended to rule out stationary black branes as well. If a black brane is stationary but not static, the Killing vector which is null on the horizon takes the form
\be\label{stationary}
\xi = \p/\p t + v \p/\p x +\Omega \p/\p \phi
\ee
where $x$ denotes some direction along the brane, i.e., a linear combination of the $x_i$, and $\phi$ denotes a rotation in the transverse space. One can first perform a boost in the $(t,x)$ plane to a co-moving frame $(\tilde t, \tilde x)$ where the black brane is at rest. Since the flux is boost invariant, we have $F_{\tilde t \tilde x \cdots } = E$. This effectively removes the second term on the right hand side of \reef{stationary}. One can now repeat the argument above. Good coordinates near the horizon of a rotating black hole take the form $v = t + h_1(r)$, $\tilde \phi = \phi + h_2(r)$. The second coordinate transformation does not affect the flux, and the first is identical to the static case. So one again finds that if $F_{p+1}$ is regular on the future horizon, there must be a nonzero flux of energy crossing this horizon causing the black hole to grow.
QED
\vskip .5cm

The theorem holds whether or not the black brane carries a charge. A black $p$-brane can carry electric charge of a $p+2$ form, or magnetic charge of a $D-(p+2)$ form. It can also carry smeared charges of lower rank forms. All these charges produce fields which are smooth on the horizon with no flux of energy crossing it. So they do not  affect the above result. They cannot stop the black hole from growing. 

There are various extensions of this theorem. A simple one just uses Hodge duality. Consider a solution with a magnetic $q$-form field $\tilde F$ which is nonzero. If there is a function $h$ such that $F = h\ \star \tilde F$ satisfies the conditions of the theorem, then one cannot put a stationary black brane in such a solution.

A less trivial extension is the one mentioned in the introduction. Even if there are no form fields $F$ in the higher dimensional theory, there can be solutions which, after dimensional reduction,  have a Maxwell field satisfying the conditions of the theorem. One cannot add stationary black branes to such a solution.  For example, consider a higher dimensional vacuum solution of the form
\be\label{KK}
ds^2 = g_{yy}(dy + Ex dt)^2 + g_{tt}dt^2 +g_{xx}dx^2 + g_{ij} dz^i dz^j
\ee
where the metric functions are independent of $t,x,y$. After Kaluza-Klein reduction on $y$, one has a Maxwell field $F_{xt} = E$, so one cannot add a stationary black string. 
We will see an example of this type of solution in Sec.~\ref{sec:fluxbranes}.

Our result certainly does not rule out the familiar planar black hole in $AdS_5 \times S^5$, even though that solution is sourced by a (self-dual) five-form. The reason is that  in the usual Poincare coordinates for $AdS_5$, the nonzero component of the flux is $F_{trx_1x_2x_3}$.  Since the radial direction is included (in which there is no translational symmetry) and a constant $r$ surface is null at the horizon, the right hand side of \reef{flux} vanishes.
 
\section{Generalization to Nonuniform Black Branes}
We now ask what happens if we relax the assumption that the black branes are translationally invariant. It is likely that static spherical black holes can exist inside these flux-branes. Indeed, exact solutions have been constructed describing a static Schwarzschild black hole inside a magnetic flux-tube  in both four  \cite{Ernst}, and higher  \cite{Ortaggio:2004kr} dimensions. In $D=4$, one can dualize the magnetic flux-tube to an electric flux-tube, but this is not possible for $D>4$. So to our knowledge  exact solutions describing  static spherical black holes in higher dimensional electric flux-branes have not yet been constructed, but are likely to exist.  

If the horizon is extended in all $p$ directions, it is likely that it cannot remain stationary even if we relax the assumption that it is uniform. We will prove this when the nonuniform directions  are compact, and then give an argument which applies to  noncompact directions. 
 
\subsection{Compact case
\label{subsec:compact}}
Suppose one direction $x_1$ is periodically identified with period $L$. It is easy to rule out stationary black branes which are inhomogeneous in this compact direction. Choose a radial coordinate $r$ such that the horizon is at constant $r$ and $x_1$ is  a coordinate along the horizon. Expanding $F_{tx_1\cdots x_p}$ in a Fourier series in $x_1$ we get
\be
F_{tx_1\cdots x_p} = F_{tx_1\cdots x_p}^{(0)} + \sum_{n\neq 0} F_{tx_1\cdots x_p}^{(n)} e^{\frac{2\pi in x_1}{L}},
\ee
The coefficients, $F_{tx_1\cdots x_p}^{(n)}$ are independent of $x_i$, but can depend on coordinates off the brane, say $r$. We can similarly expand
\be
F_{trx_2\cdots x_p} = F_{trx_2\cdots x_p}^{(0)} + \sum_{n\neq 0} F_{trx_2\cdots x_p}^{(n)} e^{\frac{2\pi in x_1}{L}},
\ee
where the coefficients $F_{trx_2\cdots x_p}^{(n)}$ again are independent of $x_i$, but can depend on $r$. The condition $dF_{p+1} = 0$ can also be expanded in a Fourier series and must hold mode by mode. In particular, if we look at just the zero mode, we get $\partial_r F_{tx_1\cdots x_p}^{(0)} = 0$. But $F_{tx_1\cdots x_p}^{(0)}$ must be nonzero at large distance from the brane since we are considering a flux brane. Since it is constant, we can now apply the argument in the uniform black brane case to conclude that stationary nonuniform black branes cannot exist.
 
This argument immediately generalizes to more than one compact direction along the brane. One can Fourier transform $F_{tx_1\cdots x_p}$ in all the compact directions. The overall zero mode must again be constant and nonzero at large distances from the black brane. 
 
\subsection{Noncompact case}
When the black brane is inhomogeneous along a noncompact direction, the argument is more subtle. One cannot just apply a Fourier transform in $x_1$ and look at the $k=0$ contribution. Since we have a constant flux at infinity, the individual $k=0$ mode diverges. In this section we will show that inhomogeneous black strings (i.e. $p=1$) cannot exist within a flux-tube  because the horizon would thin out along the string direction at a rate so rapid that the string would pinch off. We leave for future work the general $p$ case
which we believe will yield similar results.
 
\begin{figure}[htbp]
\begin{center}
\includegraphics[width=0.5\textwidth]{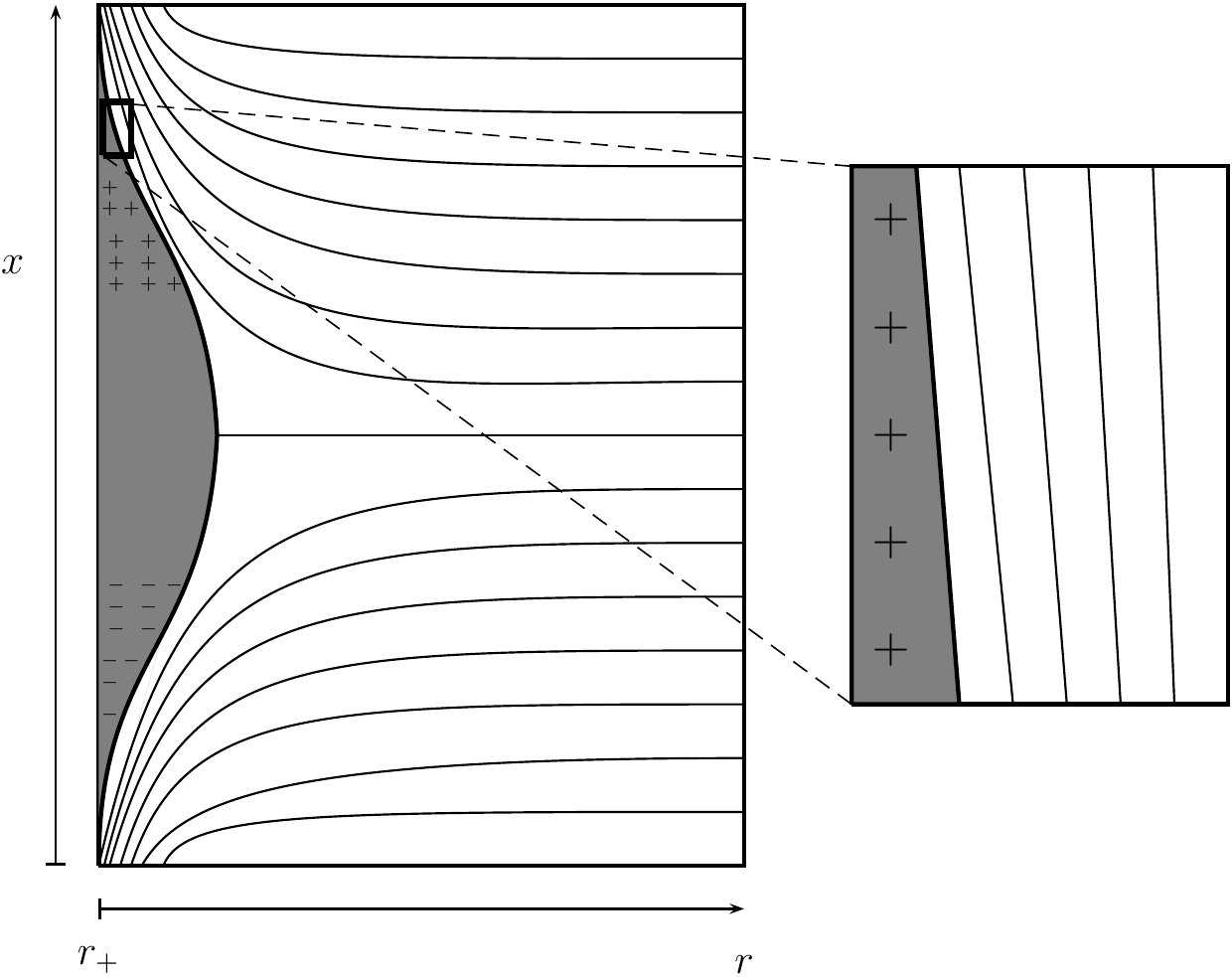}
\end{center}
\caption{Non-uniform black string in an electric flux-tube. The black lines are equipotential surfaces. The horizon is required to be an equipotential surface with value $A_t = 0$; this forces the field lines to bunch up as $x \rightarrow \pm \infty$, which means that the black string is locally positively charged at large positive values of $x$, and negatively charged at large negative values of $x$. The string remains neutral overall, however. \emph{Inset:} In the neighbourhood of some positive value of $x$, the non-uniform black string in a flux-tube should be approximated by a uniform charged black string with an electric field perturbation.}
\label{fig:cartoon}
\end{figure}

Let us suppose that the non-uniform black string in a flux-tube solution exists and consider the features it must possess. At a large radial distance $R$ away from the string, there is a uniform electric field $F_{tx} = E$. This means that asymptotically the electrostatic potential grows linearly with $x$, $A_t = E x$. In contrast, the horizon must be an equipotential surface $A_t = 0$ \footnote{This follows either from our previous argument that if $F_{tx}$ is nonzero on the horizon, there must be energy flux across the horizon, or simply from the fact that a nonzero $A_t$ on the horizon would have diverging norm.} independently of $x$, which means that the equipotential surfaces coming in from large radius cannot hit the horizon. Instead they must bunch up, producing a radial component of the electric field that grows with $x$ (see Fig.~\ref{fig:cartoon}). The radial electric field on the horizon, $F_{rt}$, will be positive for large $x>0$ but negative for large $x<0$. This requires that the black string has a positive charge density for $x>0$ and negative charge density for $x<0$, which is similar to what happens if one puts a long conducting rod in an electric field; there is charge separation with positive charge accumulating at one end and negative charge at the other.

If the horizon indeed extends to $x = \pm \infty$ without pinching off at a finite value of $x$, the $x$-dependence of the metric should be negligible near the horizon compared with the $r$-dependence, as the radial electric field grows indefinitely. Then, the geometry at large $x$ should be well described by a perturbation of a translationally invariant charged black string solution. To explore this possibility, we consider the following ansatz for the $D=d+1$ dimensional Einstein-Maxwell theory \eqref{action} with no dilaton and $p=1$:
\begin{align}\label{charged_BS}
ds^2_{d+1} &= - e^{2A} H_- H_+ dt^2 + e^{2B}\left( H_-^{-\frac{2}{d-2}} dx^2 + H_-^{-1+ \frac{2}{(d-3)(d-2)}} H_+^{-1} dr^2 \right) + e^{2C} r^2 H_-^{\frac{2}{(d-3)(d-2)}} d\Omega_{d-2}^2
, \\
& A_{\mu}dx^{\mu} = \sqrt{\frac{2 (d-1)}{(d-2)}} \left( \frac{r_-}{r_+} \right)^{\frac{d-3}{2}} H_+ e^{D}dt. \nonumber
\end{align}
Here $H_{\pm} = 1 - \left(r_{\pm}/r \right)^{d-3}$ and $A,B,C,D$ are functions of $r$ and $x$. When $A=B=C=D=0$, this is the translationally invariant electrically charged black string solution. This solution first appeared in \cite{Horowitz:2002ym} for the case $d=4$, and the general $d$ solution can be derived by dualising and uplifting the magnetically charged dilatonic black holes of \cite{Garfinkle:1990qj, Gibbons:1987ps}. The horizon topology is $S^{d-2} \times R$ and is located at $r=r_+$, while  the curvature singularity is at $r=r_-$. The extremal limit corresponds to $r_+ = r_-$ and has zero horizon area. The temperature is
\be 
T = \frac{(d-3)}{4\pi r_+} \left[ 1- \left(\frac{r_-}{r_+}\right)^{d-3} \right]^{1 - \frac{1}{(d-3)(d-2)}}.
\ee
To model the putative non-uniform black string in the neighbourhood of some large positive $x$ value,
we will start with this charged solution (with $A=B=C=D=0$) and add a small electric flux along the $x$-direction.


A  linear perturbation of the uniform charged black string can be described by:
\be
A(r,x) = \epsilon A_1(r) x, \quad B(r,x) = \epsilon B_1(r) x, \quad C(r,x) = \epsilon C_1(r) x, \quad D(r,x) = \epsilon D_1(r) x,
\ee
where $\epsilon$ is the small parameter controlling the perturbation and is proportional to the asymptotic value of the electric flux. Note that the perturbed gauge potential $A_t$ vanishes on the horizon $r=r_+$ as required by regularity. Also note that all perturbations have a linear dependence on $x$. Typically, a perturbation about a translationally invariant solution would be expanded in modes $e^{ikx}$. Here we are using the fact that we expect the dominant contribution to come from small $k$, and have kept only the linear term\footnote{We checked that the perturbation equations derived in this section are identical to those derived from a perturbation with an $e^{ikx}-$dependence once the limit $k\rightarrow 0$ is taken. Therefore the perturbation is both a gradient expansion as well as a perturbation in the amplitude of the asymptotic electric field (controlled by $\epsilon$).}.

The equations governing the perturbation come from linearizing the background Einstein and form equations \eqref{eoms}:
\begin{align}
\delta R_{ab} &= \delta F_{(a|c}F_{b)}{}^{c} - \frac{1}{2} F_{ac}F_{b d} h^{c d} - \frac{1}{2(d-1)}g_{ab} F \cdot \delta F - \frac{1}{4(d-1)} h_{ab} F^2 \\
& \qquad + \frac{1}{2(d-1)} g_{ab} F_{c e}F_{d}{}^{e} h^{c d}, \nonumber \\
& \nabla_a \left( \delta F \right)^{a b}  + \frac{1}{2} \nabla_a \left( h F^{a b} \right) - 2 \nabla_a \left( h^{c[a} F_c{}^{b]} \right) = 0.
\end{align}
There are 6 independent components of the Einstein equations and 1 independent form equation. Two equations are first order constraints: the $rx$-component Einstein equation is the momentum constraint and a linear combination of the diagonal components yields the Hamiltonian constraint. Using these constraints, the system can be shown to reduce to 4 ODE's, first order in $A_1,C_1$ and second order in $B_1,D_1$.

\begin{figure}[ht]
\centering
\includegraphics[width=1 \textwidth]{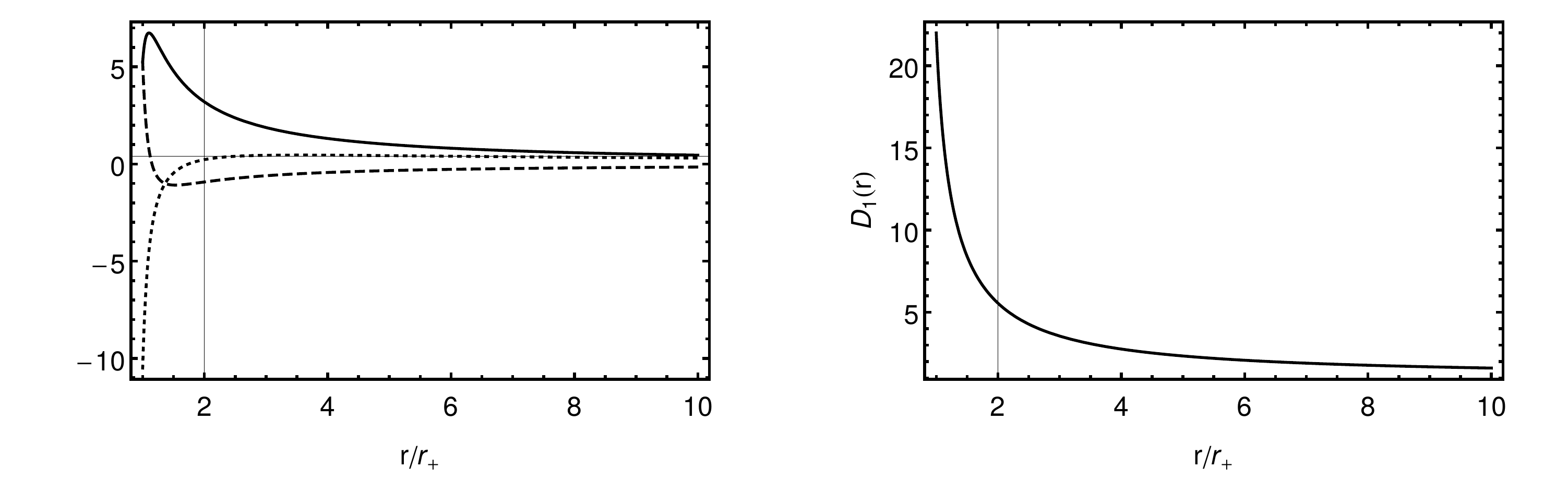}
\caption{\emph{Left Panel:} The numerical solution for $A_1$ (solid), $B_1$ (dashed), $C_1$ (dotted) for $d=4$ and $T=\sqrt{2/5}/(4\pi)$. At the horizon, $A_1(r_+) = B_1(r_+)$, as required by regularity. \emph{Right Panel:} The numerical solution for $D_1$ for the same dimension and temperature.}
\label{Fig:pertfunctions}
\end{figure}

The boundary conditions we desire are such that the perturbation is regular at the horizon and  asymptotically the perturbed metric  functions fall-off to zero and the Maxwell perturbation is that of a constant electric field whose magnitude is proportional by the expansion parameter $\epsilon$ \footnote{These boundary conditions correspond to the linearisation (in $E$) of the electric flux-brane solutions discussed later in Sec.~\ref{sec:fluxbranes}. Although those solutions are not asymptotically flat, they are when restricted to linear order in the asymptotic value of the electric field $E$, which is proportional to $\epsilon$ in the current perturbative treatment. The deviations from flatness arise at $\mathcal{O}(\epsilon^2)$.}:
\begin{align}
\label{bc:perturbation}
A_1, \, B_1, \, C_1 \to 0, \qquad D_1 \to \text{const} = 1. 
\end{align}
Requiring the perturbation to be regular at the horizon translates into the Dirichlet condition $A_1(r_+) = B_1(r_+)$ as well as additional Robin boundary conditions relating functions and their derivatives at $r=r_+$. This is a boundary value problem, and to solve it numerically we first convert to a compactified coordinate and discretize using a spectral grid. It can then be converted to a simple linear algebra problem of the form $M.v = b$, with $M$ a matrix and $b,v$ vectors. In Fig.~\ref{Fig:pertfunctions} we plot the solutions for the representative case of $d=4$, $T=\sqrt{2/5}/(4\pi)$, and $r_+ =0.632$. The solutions for other parameter values or dimensions are qualitatively similar.

It turns out that the effect of the perturbation on the horizon geometry can be understood very simply. Starting with the uniform charged solution, Eq.~\eqref{charged_BS} with $A=B=C=D=0$,  the effect of the electric field can be taken into account by promoting the black hole parameters to be functions of $x$: $r_{\pm} = r_{\pm}(x)$. To see this, note that requiring that the temperature be independent of $x$ imposes a relation between $dr_+/dx$ and $dr_-/dx$. Using this condition, one can calculate $d A_H/dx$ in terms of $d r_+/dx$,  where $A_H$ is the cross-sectional area of the horizon at fixed $x$. Returning to the linearized perturbation, one can also calculate $d A_H/dx$ in terms of $C_1(r_+)$. Equating these two expressions yields
\begin{align}
\label{C1_evolution}
\frac{dr_+}{dx} = \left( \frac{d^2-5d+5}{(d-3)(d-2)} \right) \epsilon C_1(r_+)r_+.
\end{align}
Similarly, one can calculate $d g_{xx}/dx \Big|_{r_+}$ in two different ways: first in terms of $d r_+/dx$, and then in terms of the perturbation $B_1(r_+)$. Equating the answers yields
\begin{align}
\frac{dr_+}{dx} = -\left( \frac{d^2-5d+5}{(d-3)} \right) \epsilon B_1(r_+)r_+.
\end{align} 
Thus we see that if 
\be
C_1(r_+) + (d-2) B_1(r_+) = 0
\ee
holds for our numerical solutions, then the horizon geometry is accurately modelled by the uniform charged string with the parameters $r_+$, $r_-$ promoted to functions of $x$ and subject to the constraint that $dT/dx=0$. And indeed, we find that this condition is satisfied for our solutions, up to a very small numerical error. 
Therefore, the perturbed horizon behaves exactly as the uniform charged black string made non-uniform by slowly varying parameters $r_{\pm}(x)$. 

\begin{figure}[htbp]
\begin{center}
\includegraphics[width=100mm]{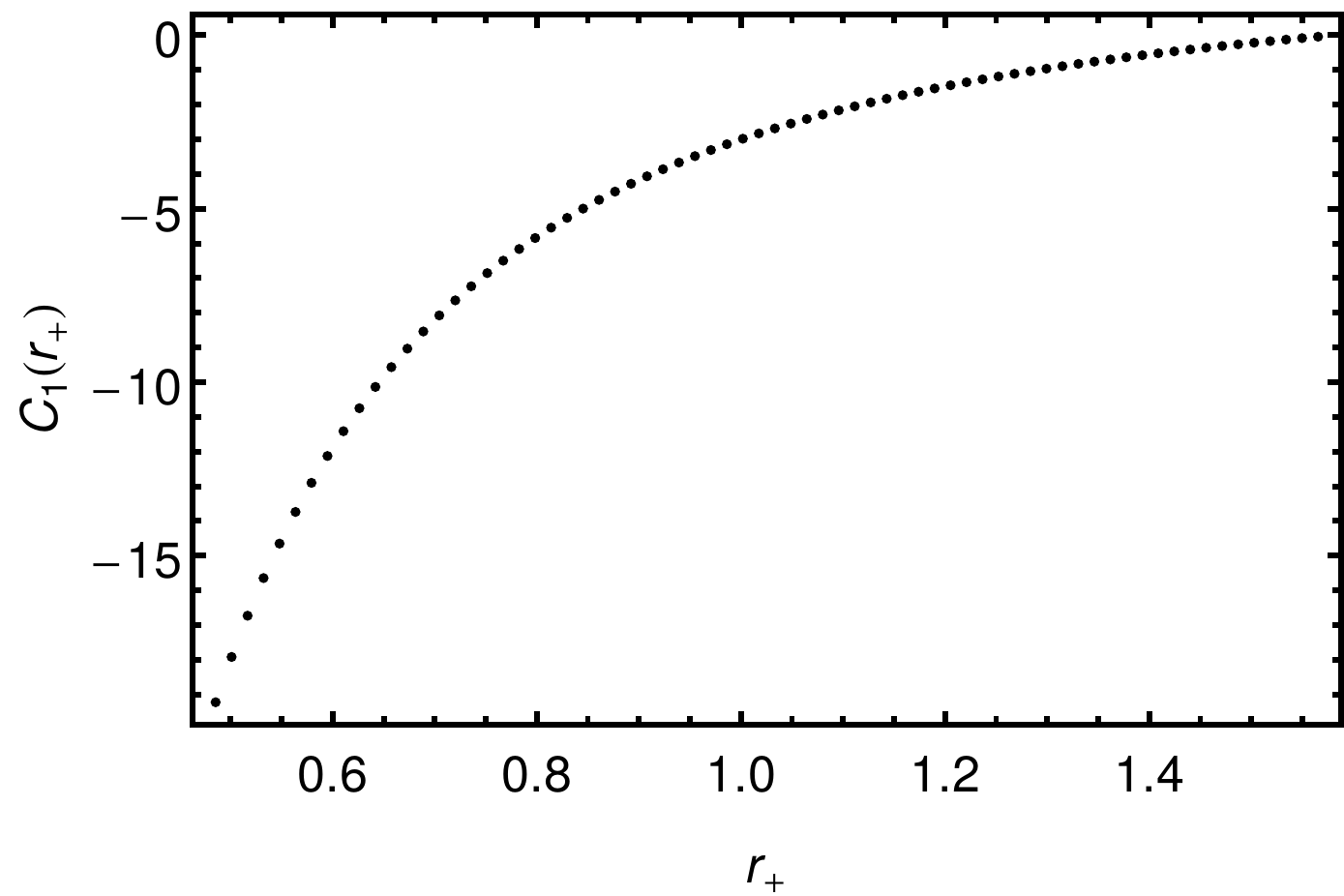}
\end{center}
\caption{$C_1(r_+)$ for $d=4$ and $T=\sqrt{2/5}/(4\pi)$. This data, along with Eq.~\eqref{C1_evolution}, shows that the black string is tapering off as $x$ increases. The apparent divergence as $r_+ \rightarrow 0$ indicates that the string will reach zero radius, i.e. pinch off, in finite $x$ distance.}
\label{fig:C1horizon}
\end{figure}

Since $C_1< 0, $
 we see that as $x \rightarrow + \infty$ the horizon thins out and the electric field evaluated at the horizon increases. It seems that there are two possibilities: either the horizon radius goes to zero in finite distance, in which case one cannot place inhomogeneous non-compact black strings in flux-tubes, or the horizon continues to shrink without pinching off as $x$ increases. This would be a new type of black hole solution, a ``spiky black hole'', that would look somewhat like Fig.~\ref{fig:cartoon} \footnote{Of course, even if this solution were found to exist, 
quantum effects would become important since the curvature is growing large as the horizon shrinks and comes closer to the singularity at $r=r_-$.}. To investigate whether such a solution is possible, let us analyse the rate at which the perturbed horizon radius decreases with $x$. In Fig.~\ref{fig:C1horizon}, we plot $C_1(r_+)$ for $d=4$ and fixed temperature, in this case  arbitrarily chosen to be $T=\sqrt{2/5}/4\pi$. We find that $C_1(r_+)$ is always negative, and appears to be diverging as $r_+ \rightarrow 0$. The divergence is well fitted by a power law,
\be \label{fit}
C_1(r_+) \sim -\alpha r_+^{-\beta} + \gamma,
\ee
with $\alpha, \beta, \gamma$ positive fit parameters that are in principle functions of $d$ and $T$. Surprisingly, we find that $\gamma = 2$ independent of $d$ or $T$, and that $\beta$ is independent of $T$ and takes on the values:
\begin{center}
\begin{tabular}{| l | l | l | l | l | l |  l | l |  l | l | l | l |}
\hline
$d$ & 4 & 5 & 6 & 7 & 8 & 9 & 10 & 11 & 12 & 13\\ 
\hline
$\beta$ & 2.000 & 1.200 & 1.091 & 1.052 & 1.034 & 1.024 & 1.018 & 1.014 & 1.011 & 1.009 \\ 
\hline
\end{tabular}
\end{center}
It is interesting to note that it appears that $\beta \rightarrow 1$ as $d \rightarrow \infty$. Lastly, $\alpha$ depends on both $d$ and $T$, but importantly is always positive. We can now use the fact that the perturbed black string is well modelled as the uniform black string with $x$-dependent parameters  by combining Eq.'s~\eqref{C1_evolution}, \eqref{fit} to find that for small $r_+$
\be
r_+(x)^{\beta} \sim c_0 - c_1 \epsilon x,
\ee
with $c_0$ a constant of integration which we take to be positive, and $c_1$ a positive constant. Clearly $r_+(x)$ pinches off at finite $x$, which is very strong evidence against the existence of these ``spiky black holes''. This result, together with the proofs of Sec.'s~\ref{sec:Theorem} and \ref{subsec:compact}, rules out any sort of extended black string in a flux-tube.

Although we only considered the case of $p=1$ for the non-uniform, non-compact case, it seems likely that the result holds for higher $p$ as well. To fully answer this question one should repeat the analysis done here. The relevant uniform charged black brane solutions can be constructed in a very similar manner to \eqref{charged_BS}, although an additional uplift would be required.

\section{Flux-branes\label{sec:fluxbranes}}
In this section we construct various flux-brane solutions to \reef{eoms}. They can be described by the general ansatz
\be\label{ansatz}
ds^2 = e^{2\alpha} ds^2_{p+1}  + e^{2\beta} dr^2 + e^{2\gamma} ds^2_{D-p-2}, \quad F_{p+1} = E \text{ vol}(ds^2_{p+1}),
\ee
where $\alpha, \beta, \gamma$ and the dilaton $\phi$ are all functions of the radial coordinate only, and $ds_{p+1}^2$, $ds^2_{D-p-2}$ are Einstein metrics obeying
\be\label{einstein}
{}^{(p+1)} R^{\mu}{}_{\nu} = \lambda_{p+1} \delta^{\mu}{}_{\nu}, \quad {}^{(D-p-2)}R^{i}{}_{j} = \lambda_{D-p-2} \delta^i{}_j ,\ee
where $\lambda_{p+1}$ and $\lambda_{D-p-2}$ are constants, indices belonging to the $(p+1)$-dimensional space are labelled with Greek letters, and the indices belonging to the $(D-p-2)$-dimensional space are labelled with lowercase Latin letters starting with $i,j,..$ Note that the form equations are already satisfied: $F_{p+1}$ is trivially closed, and although its Hodge dual in general possess a non-trivial radial dependence, it also acquires a $dr$-leg, and is therefore also closed. Also note that the gauge freedom associated with the choice of radial coordinate has not been fixed yet. We will find it convenient to work in different gauges, therefore we leave it un-fixed for now. The equations of motion for this ansatz are presented in Appendix \ref{sec:eoms}.

\subsection{Melvin Flux-branes}
The above ansatz includes both the original four-dimensional Melvin fluxtube \cite{Melvin:1963qx}, as well as its generalization to higher dimensions and non-trivial dilaton \cite{Gibbons:1987ps}\footnote{Our conventions differ from those of \cite{Gibbons:1987ps}, in  particular we are choosing to work with an electric $D-2=p+1$-form as opposed to the dual magnetic $2$-form.}. The geometry of the original Melvin solution consists of the two-dimensional Lorentz invariant worldvolume of the flux-tube, the radial direction, and a transverse circle. We will call flux-branes Melvin-like if the flux-brane has the same cohomogeneity as the original four-dimensional solution. Thus we set $p+1 = D-2$ to obtain a $p=(D-3)$-dimensional flux-brane. The solution is given by: 
\begin{align}\label{Melvin}
ds^2 &= \Lambda^{\frac{4}{a^2 (D-2)+2 (D-3)}} \left( \eta_{\mu\nu} dx^{\mu}dx^{\nu} + dr^2 \right) + \Lambda^{\frac{2 (6-2 D)}{a^2 (D-2)+2 (D-3)}} r^2 d\varphi^2, \\
& e^{2\phi} = \Lambda^{-\frac{4 a (D-2)}{a^2 (D-2)+2 (D-3)}}, \qquad F_{p+1} = E\ \text{vol}(\eta_{\mu\nu} dx^\mu dx^\nu) \nonumber,
\end{align}
where $\Lambda$ is given by
\be
\Lambda = \left(1+ \frac{a^2 (D-2)+2 (D-3)}{16 (D-2)} E^2 r^2 \right). 
\ee
Here we have imposed the  gauge $\alpha = \beta$ which is common for these solutions. For these flux-branes of cohomogeneity-2, exact solutions can be found because the $(D-p-2)$-dimensional transverse space is simply $S^1$, which has no curvature. For flux-branes of higher cohomogeneity the circle is promoted to a sphere, the curvature of which induces a coupling between the metric functions and makes the equations much harder to solve analytically \cite{Gutperle:2001mb}.

Note that the magnetic dual of \reef{Melvin} involves a two-form Maxwell field. For $p=1$ and the special value of the dilaton coupling $a  = - \sqrt{{2(D-1)/(D-2)}}$, the action \reef{action} is equivalent to pure gravity in $D+1$ dimensions (see Appendix \ref{sec:uplift}). So for this value of $a$ (in the dual magnetic frame), the magnetic dual of the above solution comes from dimensional reduction of a vacuum solution. In fact,  it can be obtained from flat Minkowski spacetime by dimensionally reducing along a combination of a translation and a rotation in a perpendicular plane \cite{Dowker:1995gb}.

Our theorem is of limited applicability for this class of flux-branes for the following reason. Near $r=0$, the backreaction of the flux-brane can be neglected and the spacetime looks flat. It is known that there are no vacuum black holes with horizon topology $\mathbb{R}^{D-3} \times S^1$, and so the stationary black holes excluded by the above theorem do not exist in the absence of flux.\footnote{It is possible that the non-standard asymptotics of the flux-brane geometries allow for these horizon topologies, i.e. that black branes of a large enough radius do exist in the flux-brane. Our theorem then implies that these black branes must not be stationary.} In order to obtain flux-brane solutions that could in principle admit small black branes at their centre, the $S^1$ must be replaced with a higher dimensional sphere. 

\subsection{Higher Cohomogeneity Flux-branes}
We now generalize the Melvin flux-branes to branes of higher cohomogeneity, so that the transverse space includes a sphere of dimension two or greater, while keeping the $p+1$-dimensional worldvolume to be Minkowski space. These solutions have been studied before \cite{Gutperle:2001mb, Saffin:2001ky, Costa:2001ifa} and include the flux-brane solutions of string theory/M-theory for appropriate choices of the parameters $(D,p,a)$:  NS flux-branes correspond to  $(D,p,a) = (10,2,-1)$ and $(10,6,1)$; the RR flux-branes correspond $(D,a) = \left(10,\frac{4-p}{2}\right)$; and  the two flux-branes of M-theory have $(D,p,a) = (11,3,0)$ or $(11,6,0)$. Here we review and extend the analysis of these solutions.

\subsubsection{Electric Flux-tubes in $D=5$}
Perhaps the conceptually simplest generalization of the above flux-brane solutions is the case of a (non-dilatonic) electric flux-tube in $D=5$. A flux-tube corresponds to $p=1$, therefore our ansatz is
\begin{align}
ds^2 &= e^{2\alpha}\left(-dt^2 + dx^2 \right) + e^{2\beta} dr^2 + e^{2\gamma} d\Omega_{2}^2, \qquad F_{2} = E dx \wedge dt, 
\end{align}
A convenient gauge choice comes from examination of the constraint equation \eqref{Grr}. No derivatives of $\beta$ appear, therefore if we impose a gauge by directly fixing $e^{2\gamma}$ to be a $\beta$-independent function (here we will use $e^{2\gamma} = r^2$), then the constraint can be used to algebraically solve for $\beta$ and the system will reduce to a single ODE for $\alpha$ (it seems that the convenience of this gauge has not been appreciated before in the literature). One can scale $E$ out of the equation since rescaling the coordinates $t,x,r$ by $\lambda$ rescales the metric by $\lambda^2$. Einstein's equation remains unchanged if we also rescale $F_2$ by $\lambda$. The resulting ODE for $\alpha$ is:
\be
3z(4e^{4\alpha}+z^2)\alpha''
+ 3(8e^{4\alpha}-5z^2)\alpha' 
- 12z(-4e^{4\alpha}+z^2)(\alpha')^2 
+ 2z^2(6e^{4\alpha}-z^2)(\alpha')^3 
- 4z = 0,
\ee
where $z \equiv E r$. 

We were unable to determine the  general solution analytically, but a simple closed form solution does exist, and is
\be 
e^{2\alpha} = \sqrt{\frac{5}{6}} E r, \qquad e^{2\beta} = \frac{5}{2}, \qquad e^{2\gamma} = r^2. 
\ee
This solution is clearly singular at $r=0$ and hence is not of physical interest. Nevertheless, it is quite useful since it can be shown that this solution is an attractor solution for the large-$r$ behaviour of general solutions \cite{Gutperle:2001mb}. To see this, linearise around the exact singular solution, $\alpha = \alpha_{\text{singular}} + \delta \alpha$. The solution to the linearised perturbation takes the form
\be 
\delta \alpha  = r^{-1} \left[ c_1 \sin(2 \ln r) + c_2 \cos(2 \ln r) \right],
\ee
where $c_{1,2}$ are constants of integration. For large $r$, the perturbation oscillates with a decaying envelope and asymptotes to the above exact singular solution, indicating that it is indeed an attractor.

\begin{figure}[ht]
\centering
\includegraphics[width=0.8 \textwidth]{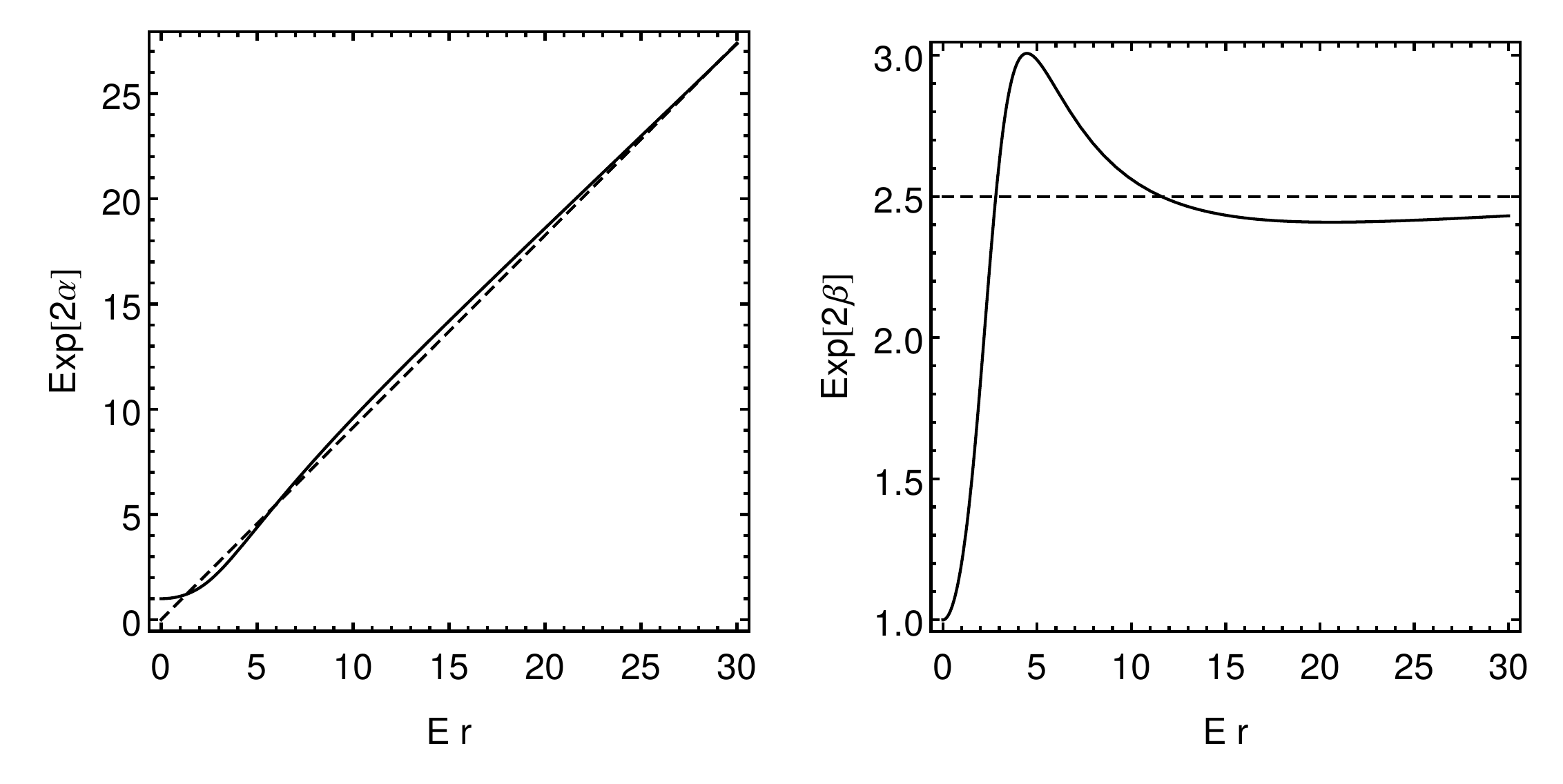}
\caption{The metric functions $e^{2\alpha}$ (left panel) and $e^{2\beta}$ (right panel) for a non-dilatonic electric flux-tube in five dimensions ($D=5$, $p=1$, $a=0$). The solid curves correspond to the numerical, non-singular solution, and the dotted curve corresponds to the singular attractor solution.}
\label{Fig:numerics}
\end{figure}

The solution we are interested in is regular along the axis $r=0$. We did not succeed in finding it analytically, but the equations can be solved numerically once the boundary conditions are supplied. The large-$r$ behaviour is that of the attractor, and the small $r$ behaviour is simply regularity at the origin. There will be a curvature singularity unless $e^{2\alpha} \rightarrow \text{const}$, and $e^{2\beta} \rightarrow 1$. Expanding the equations in a power series around $r=0$ yields
\be
e^{2\alpha} = 1+\frac{z^2}{9}+\frac{2 z^4}{405} + O\left(z^6\right), \qquad e^{2\beta} = 1 + \frac{7 z^2}{36}+\frac{61 z^4}{6480} + O\left(z^6\right)
\ee
We have used the scaling symmetry of the Minkowski directions to set an overall constant to 1. The boundary condition required by regularity is then $A(0) = 1$, $A'(0) = 0$, where $A = e^{\alpha}$. In Fig.~\ref{Fig:numerics} we plot the numerical solution for this flux-tube. From the plot, the non-singular solution can be seen to asymptote to the attractor solution at large $r$. 

This electric flux-tube is much better behaved asymptotically than the Melvin flux-branes. One can see from \reef{Melvin} that in the Melvin case, the $S^1$ in the transverse space shrinks to zero size as $r \to \infty$. In the current solution, the effect of the flux-tube on the asymptotic geometry is weaker, and the $S^2$ in the transverse space continues to grow asymptotically. However, the solution is still not asymptotically flat in the transverse directions.  It is asymptotically a cone. 

Applying the results of Sections 2 and 3 to this solution, we see that we cannot put a stationary black string inside the electric flux-tube. Stated intuitively, there is no way to  prevent stress energy from flowing into the black string, which would cause it to grow.

There is a simple physical interpretation of this result in terms of a stretched horizon\footnote{We thank Juan Maldacena for suggesting this.}. The stretched horizon has finite electrical conductivity, so if one puts a black string along an electric field, a current will flow. Since the resistance is nonzero, the current will generate heat and cause the black string to grow. So a static black string is impossible. This is very similar to the discussion in \cite{Horowitz:2013mia} where an exact solution was found showing a black hole growing when an electric field is applied. 
 
\subsubsection{Kaluza-Klein flux-tube in $D=5$}
An interesting extension of this five-dimensional example is to include a dilaton, and consider it in the context of Kaluza-Klein theory.  As before, we choose to impose the gauge $e^{2\gamma} = r^2$, and eliminate $\beta$ using the constraint. Now the system has been reduced to two coupled ODE's in $\alpha$ and $\phi$. As noted in Ref.~\cite{Gutperle:2001mb}, the  dilaton equation of motion becomes a copy of the $\alpha$ equation of motion if we set
\be
\phi = - \frac{3a}{2} \alpha.
\ee
Therefore the system has again been reduced to a single ODE, and the inclusion of the dilaton comes at no cost in complexity. We omit the presentation of the equation as it is rather lengthy. Once again, we find that the general solution must be solved for numerically, and that a simple, singular solution exists.

For the case $a = a_{\text{KK}} = - \sqrt{8/3}$, the flux-tube can be uplifted to a vacuum solution of Einstein gravity (see Appendix \ref{sec:uplift}). The flux-tube has been geometrized, and the six-dimensional line element is now
\be\label{KKupliftoffluxtube}
ds_6^2 = e^{-3\alpha}\left(dy + E x dt \right)^2 + e^{3\alpha} \left( -dt^2 + dx^2 \right) + e^{\alpha + 2\beta} dr^2 + e^{\alpha + 2\gamma} d\Omega_2^2.
\ee
where we have chosen the gauge $A =  E x dt$. While the metric \eqref{KKupliftoffluxtube} has  three translational symmetries, they do not all commute. The Killing fields include $\p/\p t, \p/\p y$, and $\p/\p x - Et \p/\p y$. This is a timelike version of the Bianchi Type II symmetries of a homogeneous anisotropic cosmology. To gain some insight into this Ricci-flat solution, we can examine its large-$r$ behaviour, which is governed by the singular attractor solution:
\be
e^{2\alpha} = \left(\frac{3}{2}\right)^{1/4} (E r)^{1/2}, \qquad e^{2\beta} =  \frac{27}{16}, \qquad e^{2\gamma} = r^2.
\ee
Note that the KK circle is pinching off at infinity. 

Since one cannot put a stationary black string in the five-dimensional flux-tube, one cannot put a stationary black two-brane in this vacuum solution. One can see this directly in six-dimensions, by deriving the following contradiction. Adding a stationary uniform black two-brane would produce a metric of the following form:
\be
ds^2_6 = g_{yy}\left(dy + E x dv \right)^2 + h\left[-f dv^2 + 2dvdr \right ] + g_{xx} dx^2 +r^2 d\Omega
\ee
where $v$ is an ingoing null coordinate, and $f$ vanishes at some radius $r_0$ denoting the horizon. The metric functions $g_{xx}, g_{yy}, h$ depend only on $r$. The vector $\ell = \p/\p v - Ex \p/\p y$ is null on the horizon but it is not a Killing vector. So the six-dimensional spacetime does not have a Killing horizon. One can show that the vector $\ell$ is tangent to a null geodesic, and has zero expansion, but nonzero shear.  This violates  the Raychaudhuri equation \reef{Raychaudhuri} and shows that the assumption of a stationary solution is inconsistent with the field equations. To see why the shear is nonzero, consider a small bundle of light rays extended in the $x$ and $y$ directions with thickness $\Delta x, \Delta y$. Starting at $v = v_0$, this bundle has a rectangular cross-section. Under evolution by $\ell$, the rectangle gets distorted with  geodesics pushed forward in the $y$ direction by an amount proportional to $x$. This produces shear. 

This argument is a Lorentzian version of the one  given by Iizuka et al  \cite{Iizuka:2014iva} who consider five dimensional black holes having Bianchi symmetry on the horizon. They  show that for Bianchi types II, VI$_0$, or VII$_0$, if $\p/\p v - \ell$ is not a Killing field, then the shear will be nonzero and the solution cannot be stationary.

\subsubsection{General Case}
Having studied five-dimensional flux-tubes we now discuss the general case. Much of the above analysis carries over in higher dimensions. The gauge $e^{2\gamma} = r^2$ is imposed, and the constraint equation is used to algebraically solve for $\beta$. The system now only involves two coupled and undetermined functions, $\alpha$ and $\phi$. As in the five-dimensional dilatonic flux-brane, the $\phi$ equation of motion becomes identical to the $\alpha$ equation of motion after a rescaling \cite{Gutperle:2001mb} \footnote{As Ref.~\cite{Gutperle:2001mb} worked with the Hodge dual picture, their expression is related to ours, \eqref{integralofmotion}, via $p+1 \rightarrow 10 - (p+1)$.}
\be\label{integralofmotion}
\phi = -\frac{a(D-2)}{(D-p-2)} \alpha.
\ee
Therefore the equations of motion for this general ansatz have been reduced to a single ODE, the presentation of which we omit as it is quite long. As before a simple analytic family of singular solutions exists:
\begin{align} 
e^{2\alpha} &= \left[ \frac{(2+a^2)(D-2) + 2p(D-2) - 2p^2}{4(D-2)(D-p-3)} \left( E r \right)^2 \right]^{\frac{2(D-p-2)}{a^2(D-2) + 2(1+p)(D-p-2)}} , \\ \nonumber \\
e^{2\beta} &= \frac{\left(a^2 (D-2) (D-p-3)+2 (p+1) (D-p-2)^2 \right) \left(a^2 (D-2)+2 \left(D(p+1) - (p+1)^2 -1 \right)\right)}{(D-p-3) \left(a^2 (D-2)+2 (p+1) (D-p-2)\right)^2}. 
\end{align}
These solutions have been noticed by numerous authors for various values of the parameters $(D,p,a)$ \cite{Gutperle:2001mb, Saffin:2001ky, Costa:2001ifa}. Although  clearly singular at $r=0$, these exact solutions are still quite useful as they are often attractors for the large-$r$ behaviour of more general solutions \cite{Gutperle:2001mb}. As before, this can be determined by linearising around the exact solution, $\alpha = \alpha_{\text{singular}} + \delta \alpha$. The solution to the linearised perturbation takes the form
\be 
\delta \alpha  = r^{-q} \left[ c_1 \sin(\nu \ln r) + c_2 \cos(\nu \ln r) \right],
\ee
where $c_{1,2}$ are constants of integration, and $q,\nu$ are constants depending on $(D,p,a)$. If both $q, \nu^2 >0$, then the perturbation oscillates with a decaying envelope and asymptotes to the above exact singular solution for large $r$, indicating that it is an attractor. We have checked the singular solution is an attractor for the following solutions: the two 5d flux-tubes presented above, and all of the flux-branes of string theory/M-theory with co-dimension greater than two.

The solutions we are interested in  are, of course, nonsingular at the origin. We did not succeed in finding analytic non-singular solutions, but the equations can be solved numerically, just as they were in the $D=5$ cases considered above. We therefore see that there is a large class of non-singular flux-brane solutions for which our theorem applies, including flux-branes that appear in string/M-theory. These flux-branes cannot be ``blackened'' without introducing  time-dependence\footnote{Here a comment is in order: The flux-brane solutions are qualitatively different from the $p$-brane solutions of string theory. The $p$-branes are singular supergravity solutions that possess degrees of freedom associated with their worldvolume, whereas the flux-branes are completely non-singular (the simple family of singular solutions is not physical), and have no worldvolume degrees of freedom. Therefore, the inability to ``blacken'' the flux-branes carries no implications for a worldvolume theory.}. 

\section{Discussion}
We have shown that one cannot put stationary black branes inside flux-branes. This was rigorously established for uniform black branes or black branes that are nonuniform in compact directions. For the case of black strings inside a flux-tube, we have given numerical evidence that this result extends to the noncompact case, and we expect that it extends to all cases where the horizon is noncompact. 

A translationally invariant black string (or black brane) is subject to Gregory-Laflamme instabilities. One can stabilize it by compactifying the direction it is extended along. Our result shows that even when it is stable, the black string cannot remain stationary inside a flux tube.
 
An interesting open question in this compactified context is the following\footnote{We thank Henriette Elvang for raising this question.}. In vacuum gravity with one direction compactified on a circle, there is a well studied black hole - black string transition in the space of static solutions (see \cite{Headrick:2009pv} and references therein). One can start with a small spherical black hole and increase its size. One obtains a continuous family of solutions in which the spherical black hole gets distorted when it approaches the size of the circle. It then  makes a transition to a nonuniform black string and eventually turns into a uniform black string.  Now suppose we start with an electric flux-tube wrapping the compact direction. As we argued in section 2, a small static spherical black hole should still exist. We can continuously increase the size of this  black hole and it is likely that there will again be a transition to a nonuniform black string. But we have seen that the nonuniform black string cannot be static. So when does the static solution stop existing? 
The intuitive picture of the electric field inducing currents on the stretched horizon given in section 4.2.1 suggests that it will not be until  the nonuniform black string forms.

What is the analog of this intuitive picture for higher rank forms? Can one understand the absence of static black two-branes inside $F_3$ flux vacua, by postulating that the stretched horizon contains strings which move when placed inside this flux? This seems likely since one can dimensionally reduce a flux $p$-brane along $p-1$ directions to obtain a flux-tube. We have argued that one cannot put a static black string in this flux-tube due to the existence of currents on the stretched horizon which imply the existence of charged particles. From the higher dimensional standpoint, these charged particles are charged $p-1$ branes. We should emphasize that there is no actual charged matter in the spacetime. These charged objects arise in an effective description of form fields interacting with black holes. This property of stretched horizons deserves further investigation.

\vskip 0.5cm
\centerline{\bf Acknowledgements}
\vskip 1cm
It is a pleasure to thank H. Elvang, V. Hubeny, J. Maldacena, and A. Puhm for discussions. This work was supported in part by NSF grant PHY12-05500, by NSF grant PHYS-1066293 and the hospitality of the Aspen Center for Physics. It was also supported in part by JSPS KAKENHI Grants No. 26400280.

\begin{appendix}
\section{Equations of Motion}
{\label{sec:eoms}}
Here we present the equations of motion \reef{eoms} for the ansatz \eqref{ansatz}. The form equations are already satisfied as noted above. The three independent (trace reversed) Einstein equations are
\begin{align}
R^{\mu}{}_{\nu} &= \left[ - \left[\alpha''-\alpha'\beta' + (p+1)(\alpha')^2 + m \alpha' \gamma' \right]e^{-2\beta} + \lambda_{p+1}e^{-2\alpha} \right] \delta^{\mu}_{\nu},\\
& = \tau^{\mu}{}_{\nu} =  -\frac{m}{2(D-2)} E^2 e^{a\phi -2(p+1)\alpha} \delta^{\mu}{}_{\nu} \nonumber.
\end{align}
\begin{align} 
R^r{}_r &= \left[-(p+1) \alpha'' - m\gamma'' + (p+1)\alpha' \beta' + m \beta' \gamma' -(p+1)(\alpha')^2 -m(\gamma')^2 \right]e^{-2\beta} \\
& = \tau^r{}_r = \frac{1}{2}e^{-2\beta}(\phi')^2 + \frac{p}{2(D-2)} E^2 e^{a\phi -2(p+1)\alpha}. \nonumber 
\end{align}
\begin{samepage}
\begin{align} 
R^i{}_j &= \left[ -\left[\gamma'' -\beta' \gamma' + m (\gamma')^2 + (p+1) \alpha' \gamma' \right]e^{-2\beta}  + \lambda_m e^{-2\gamma} \right] \delta^i{}_j \\
& = \tau^i{}_j = \frac{p}{2(D-2)}E^2 e^{a\phi - 2(p+1)\alpha} \delta^i{}_j \nonumber. 
\end{align}
\end{samepage}
where $m=D-p-2$ and primes indicate derivatives with respect to $r$. A very useful linear combination of these equations is the ``Hamiltonian constraint" 
\begin{samepage}
\begin{align} \label{Grr}
G^r{}_r &= e^{-2\beta}\left[\frac{1}{2}m(m-1)(\gamma')^2 + \frac{1}{2}p(p+1)(\alpha')^2 + m(p+1)\alpha'\gamma'\right] \\
& \qquad \qquad - \frac{m}{2} \lambda_m e^{-2\gamma} - \frac{1}{2}(p+1)\lambda_{p+1} e^{-2\alpha} \nonumber \\
&= T^r{}_r = \frac{1}{4}(\phi')^2 e^{-2\beta} + \frac{E^2}{4}e^{a\phi - 2(p+1)\alpha} \nonumber.
\end{align}
\end{samepage}
Lastly, the dilaton equation of motion is
\be\label{dilaton} e^{-(p+1)\alpha-\beta-m\gamma} \partial_r \left( e^{(p+1)\alpha - \beta + m\gamma} \partial_r \phi \right) + \frac{a E^2}{2} e^{a\phi - 2(p+1)\alpha} = 0. \ee

\section{Uplift of Kaluza-Klein Electric Flux-tubes}
{\label{sec:uplift}}
If the dilatonic coupling in \eqref{action} takes the special value 
\be
a = a_{\text{KK}} = - \sqrt{\frac{2(D-1)}{(D-2)}}, 
\ee
and $p=1$ (so $F_{p+1}$ is a standard Maxwell field)
then a solution of the Einstein-Maxwell-dilaton theory \eqref{action} in $D$ dimensions uplifts to a solution of the vacuum Einstein equations in $D+1$ dimensions. The uplifted line element is
\be
ds_{D+1}^2 = e^{-\sqrt{\frac{2(D-2)}{(D-1)}}\phi} \left(dy + A_{\mu}dx^{\mu} \right)^2 + e^{\sqrt{\frac{2}{(D-2)(D-1)}} \phi} ds_{D}^2
\ee
Here $A$ is the 1-form gauge potential related to the field strength via $F=dA$.

\end{appendix}

\vskip 2cm
\bibliography{refs}{}
\bibliographystyle{utphys}

\end{document}